\documentclass[letterpaper]{article} 
\usepackage{aaai25}  
\usepackage{times}  
\usepackage{helvet}  
\usepackage{courier}  
\usepackage[hyphens]{url}  
\usepackage{graphicx} 
\usepackage{natbib}  
\usepackage{caption} 
\usepackage{algorithm}
\usepackage{algorithmic}

\usepackage{newfloat}
\usepackage{listings}

\usepackage{latexsym}
\usepackage{amssymb}
\usepackage{amsmath}
\usepackage{amsthm}
\usepackage{booktabs}
\usepackage{enumitem}
\usepackage{color}
\usepackage{multirow} 
\usepackage{pgfplots}
\usepackage{subfigure}
\usepackage{pifont}

\DeclareCaptionStyle{ruled}{labelfont=normalfont,labelsep=colon,strut=off} 
\floatstyle{ruled}
\newfloat{listing}{tb}{lst}{}
\floatname{listing}{Listing}
\title{SongSong: A Time Phonograph for Chinese SongCi Music from Thousand of Years Away}
\author{
    Jiajia Li\textsuperscript{\rm 3,\footnotemark[2]}, 
    Jiliang Hu\textsuperscript{\rm 1,\footnotemark[2]},
    Ziyi Pan\textsuperscript{\rm 1}, 
    Chong Chen\textsuperscript{\rm 4}, 
    Zuchao Li\textsuperscript{\rm 2,\footnotemark[1]}, 
    Ping Wang\textsuperscript{\rm 3,\footnotemark[1]}, 
    Lefei Zhang\textsuperscript{\rm 2}
}
\affiliations{
    \textsuperscript{\rm 1}Key Laboratory of Aerospace Information Security and Trusted Computing, Ministry of Education, \\School of Cyber Science and Engineering, Wuhan University, Wuhan, China, \\
    \textsuperscript{\rm 2}School of Computer Science, Wuhan University, Wuhan, China, \\
    \textsuperscript{\rm 3}School of Information Management, Wuhan University, Wuhan, China,\\
    \textsuperscript{\rm 4}ShenYang Conservatory Of Music, ShenYang, China.
}

\usepackage{bibentry}

\begin{document}

\maketitle

\begin{abstract}
Recently, there have been significant advancements in music generation. However, existing models primarily focus on creating modern pop songs, making it challenging to produce ancient music with distinct rhythms and styles, such as ancient Chinese SongCi. In this paper, we introduce SongSong, the first music generation model capable of restoring Chinese SongCi to our knowledge. Our model first predicts the melody from the input SongCi, then separately generates the singing voice and accompaniment based on that melody, and finally combines all elements to create the final piece of music. Additionally, to address the lack of ancient music datasets, we create OpenSongSong, a comprehensive dataset of ancient Chinese SongCi music, featuring 29.9 hours of compositions by various renowned SongCi music masters. To assess SongSong's proficiency in performing SongCi, we randomly select 85 SongCi sentences that were not part of the training set for evaluation against SongSong and music generation platforms such as Suno and SkyMusic. The subjective and objective outcomes indicate that our proposed model achieves leading performance in generating high-quality SongCi music.
\end{abstract}


\renewcommand{\thefootnote}{\fnsymbol{footnote}}
\footnotetext[2]{These authors contribute equally to this work.}
\footnotetext[1]{Corresponding authors.}

\renewcommand {\thefootnote} {\arabic {footnote}} 


%
\begin{links}
    \link{Project}{https://zcli-charlie.github.io/songsong/}
\end{links}



\section{Introduction}
\begin{figure}[htbp]
	\centering 
	\includegraphics[scale=0.3]{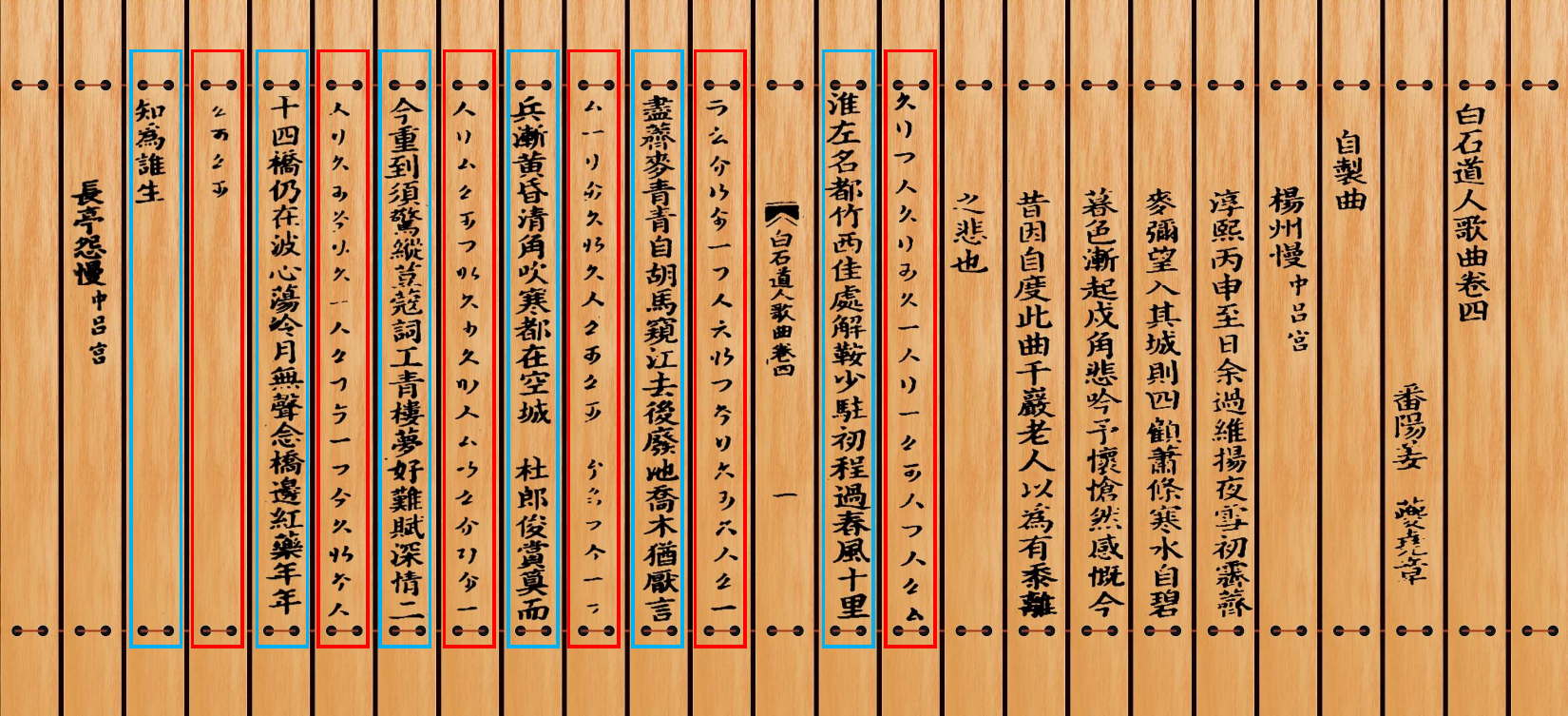} 
	\caption{A piece of ancient Chinese SongCi music recorded using Chinese Gongche notation, which is vertically arranged and uses special Chinese symbols as musical notes. The blue box indicates the SongCi, and the red box indicates the notes.} 
    \label{bg}
\end{figure}

SongCi, as a brilliant pearl in the history of Chinese poetry, is an important carrier of excellent Chinese culture and carries the spiritual pursuit of the Chinese nation for thousands of years. As shown in Figure \ref{bg}, these ancient poems that have been passed down to today can not only be recited, but also sung as lyrics. Unfortunately, due to historical changes and the loss of ancient music scores, the original musical forms of SongCi have largely disappeared and buried in the torrent of history.

In recent years, music understanding \citep{li2024music,tian2024n} and generation have seen considerable advancements. In terms of music generation, numerous studies have focused on generating singing voices \citep{liu2022diffsinger,chen2020hifisinger} and melodies \citep{yu2021conditional,zhang2022relyme,lv2022re}. Deep learning techniques can potentially restore SongCi music, but there is presently no open-source dataset available for ancient Chinese SongCi. Existing Chinese music datasets, like M4Singer \citep{zhang2022m4singer} and OpenCPOP \citep{wang2022opencpop}, are limited to pop songs. Currently, most models capable of generating complete music with singing and accompaniment are commercial products, such as Suno and SkyMusic. These models leverage the powerful generative abilities of large language models (LLMs) \citep{zhao2023survey} to create full audio based on user-provided text prompts and lyrics. Additionally, users can input audio as a style reference for the model to generate music accordingly. However, the generation process in these models is often poorly controllable and highly random. Furthermore, the training data for these models primarily consists of contemporary popular songs, resulting in generated music that predominantly reflects mainstream styles instead of ancient music styles. Although \citet{shan2023lingge} previously proposed an ancient Chinese poem-to-song system, due to the insufficiency of structure superiority and the lack of ancient music corpus, the music generated by this system is still closer to modern music. 


In this study, we introduce SongSong, a music generation model that can perform Chinese SongCi. The model utilizes the auto-regressive Transformer and the diffusion technology. It takes a SongCi poem as input, first creating a melody that corresponds to the poem. Next, it synthesizes the singing voice and generates accompaniment based on the music score derived from the melody. Finally, the singing voice, melody, and accompaniment are combined to create the final piece of music. To tackle the scarcity of ancient music datasets and aid in the restoration of ancient music, we present OpenSongSong, a comprehensive dataset of ancient Chinese SongCi music, totaling 29.9 hours of SongCi music. We conduct experiments comparing SongSong with Suno and SkyMusic, evaluating the results from both subjective and objective viewpoints. The findings indicate that our proposed model can generate high-quality SongCi music that aligns with the traditional style of SongCi. The key contributions of this work can be summarized as follows:

\begin{enumerate}
	\item[(1)]We propose SongSong, the first music generation model that can perform SongCi music to our knowledge.
    \item[(2)]We have developed a comprehensive SongCi music dataset to address the shortage of publicly available ancient Chinese music datasets and to support the restoration of SongCi music.
    \item[(3)]Experimental results demonstrate that the proposed model achieves state-of-the-art performance in generating SongCi music.
\end{enumerate}

\section{Related Work}
\textbf{Singing Voice Synthesis} Singing Voice Synthesis (SVS) is an attracting technique that uses musical score information (such as lyrics, tempo, pitch, etc.) to generate natural and expressive singing voices \citep{yamamoto2020parallel}. In recent years, deep learning has brought revolutionary progress in the field of SVS, achieving significant improvements over traditional methods. Early deep learning-based systems use Feedforward Neural Networks (FFNN) \citep{nishimura2016singing}, which outperforms HMM-based \citep{rabiner1989tutorial} systems by predicting acoustic features directly from musical scores. Further developments introduce Long Short-Term Memory (LSTM) \citep{kim2018korean} and Convolutional Neural Networks (CNNs) \citep{nakamura2019singing,nakamura2020fast}, which enhances the ability to model long-term dependencies and acoustic features in singing voices. Generative Adversarial Networks (GAN) \citep{hono2019singing,chandna2019wgansing} is also integrated into SVS systems to mitigate the over-smoothing problem, leading to more natural and expressive singing voice synthesis. Moreover, state-of-the-art deep learning architectures such as Transformer-based models \citep{vaswani2017attention} like XiaoiceSingv \citep{lu2020xiaoicesing}, GAN-based models like HifiSinger \citep{chen2020hifisinger}, and diffusion-based models \citep{ho2020denoising} like DiffSinger \citep{liu2022diffsinger}, have further improved the quality of SVS. 



\noindent{\textbf{{Melody Generation}} Lyric-to-melody generation is a key task in automatic composition, which involves generating a melody that matches a given lyric. It usually uses an end-to-end (E2E) model to generate melodies directly from lyrics. \citet{bao2019neural} and \citet{yu2021conditional} use sequence-to-sequence models to generate melodies from lyrics. However, these E2E models require a large amount of paired lyrics and melody data, and obtaining a sufficient amount of paired data is both difficult and costly. To address this problem, \citet{sheng2021songmass} avoid the reliance on paired data by training lyric-to-lyric and melody-to-melody models separately and interacting them in subsequent stages. However, since unpaired data is not fully utilized in learning the correlation between lyrics and melody, the consistency of lyrics and melody features cannot be ensured. \citet{ju2022telemelody} propose TeleMelody, a generative model with two-stage: lyrics-to-template and template-to-melody. The template bridges the gap between lyrics and melody, promotes better alignment of features, and improves the controllability of generated melody.

\section{Method}

Figure \ref{framework} illustrates the structure of our model, SongSong, which is composed of four modules. It takes SongCi poetry as input and produces SongCi music audio as output. In the first stage, the lyric-to-rhythm module predicts the rhythm, which includes specific tonality and chord details, for each word in the poetry. In the second stage, the rhythm is processed by the rhythm-to-melody module to determine the corresponding note for each word. Next, we transform the sequence of notes into a melody MIDI file along with a config file. This config file contains information about the phonemes of the lyrics, the notes, and the fundamental frequency (f0) for the audio to be created, which is used to generate the singing voice audio. The melody audio is then utilized by the accompaniment generation module to create accompaniment played by other instruments, enhancing the overall richness of the audio. Ultimately, the complete SongCi music audio is produced by merging the singing voice audio, melody audio, and accompaniment audio.

\subsection{Lyric To Melody}
\begin{figure}[htbp]
	\centering 
	\includegraphics[scale=0.435]{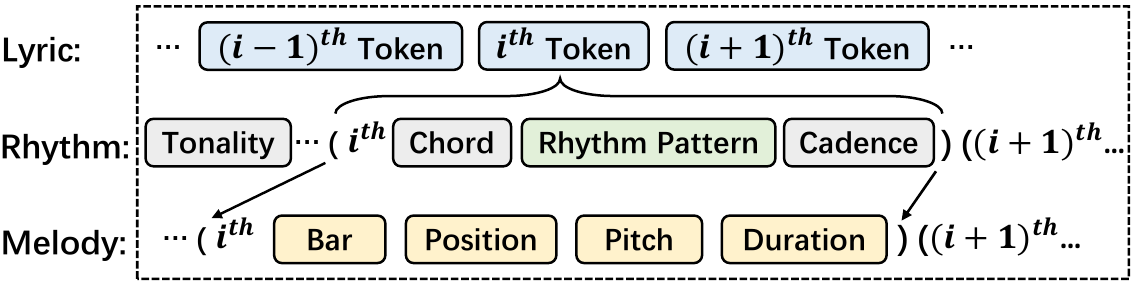} 
	\caption{The relationship between lyric, rhythm, and melody.} 
    \label{melody}
\end{figure}

The differences between lyrics and melody are quite pronounced. To create a model that can directly link lyrics to melody, a substantial amount of training data is necessary. However, since there is currently no extensive dataset for Chinese SongCi music, we utilize a two-stage method involving rhythm transitions to transform lyrics into melody. 

\begin{figure*}[htbp]
	\centering 
	\includegraphics[scale=0.64]{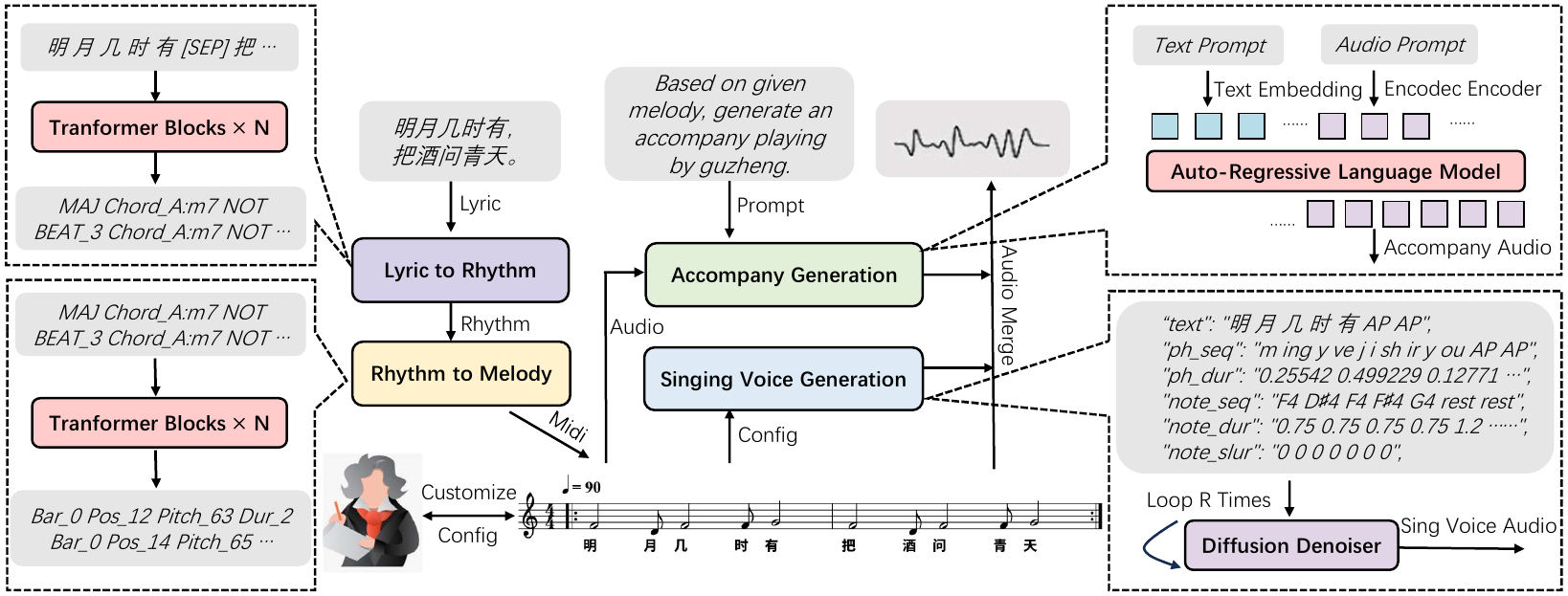} 
	\caption{The structure of our proposed model, SongSong. The English meaning of input SongCi is ``How rare the moon,so round and clear! With cup in hand,I ask of the blue sky."}
    \label{framework}
\end{figure*}

Figure \ref{melody} illustrates the connections between lyrics, rhythm, and melody. The format of rhythm sequence is inspired by \citet{ju2022telemelody} and encompasses tonality, chords, rhythm patterns, and cadences. Tonality consists of a scale and a root note, with each piece having only one tonic, meaning each rhythm also contains just one tonality. However, a rhythm can include multiple chords, rhythm patterns, and cadences corresponding to the lyrics. Chords are collections of sounds with specific interval relationships, determined by a chosen chord progression. Cadences indicate the conclusion of a melodic section and can be classified as "no cadence," "half cadence," or "authentic cadence," depending on whether the associated lyric is a regular token, a pause token, or a final pause token. The rhythm pattern is derived from the lyric-to-rhythm module which is consist of stacked Transformer blocks, essentially representing the beat that aligns with the lyrics. Let $X$ denote the lyric sequence, which has a length of $n$, and $Y$ represent the beat sequence, which is of the same length.
$$
Y=\textrm{Softmax}(\textrm{Transformer}(X))
$$

The rhythm-to-melody module has the same structure as lyric-to-rhythm module. It takes the rhythm as input and produces a note sequence that corresponds to the lyrics. Each note is represented as a quadruple that includes the bar, position, pitch, and duration. The bar and position are used to determine the initial sound time of the note, pitch indicates the note's pitch, and duration specifies how long the note sounds. Let $Z$ represent the note sequence, which has a length of $n$.

$$
Z=\textrm{Softmax}(\textrm{Transformer}(Y))
$$

The training goal of these two modules is to minimize the negative log-likelihood on the lyric-to-rhythm data $(X_i Y_i)$ or the rhythm-to-melody data $(Y_i,Z_i)$, which serves as the loss function. For instance, the optimization goal for the lyric-to-rhythm module is as follows.

$$
\mathcal L_{lt}=-\mathbb{E}_{(X,Y)}logP(Y_i|X_i)
$$


\subsection{Melody To Singing Voice}

Our singing voice generation model utilizes phonetic units, meaning the input consists of a phonetic sequence $P$ instead of a lyric sequence $X$. The length of $P$ corresponds to the total number of phonetic units $m$. However, having just $P$ is insufficient; we also need to determine the duration of each phoneme, represented as $Q$, which also has a length of $m$ and cannot be calculated using rule-based methods. Furthermore, the singing voice module requires melody information to ensure the lyrics are sung at the correct pitch. In the previous section, we derived a sequence of note quadruplets $Y$, which includes the note sequence $U$ and the note duration sequence $V$. These are matched one-to-one with each lyric, resulting in both $U$ and $V$ having a length of $n$. To produce a complete song, we also need the fundamental frequency information $F0$, which has a length equal to the number of frames $T$ in the generated audio. $F0$ must also be inferred using a machine learning model. Figure \ref{singvoice} illustrates the detailed design of the singing voice generation module. The variance encoder is a speech acoustic encoder based on FastSpeech2 \citep{ren2020fastspeech}, consisting of Transformer layers. It takes $P$ as input and produces the phonetic acoustic feature $H^{VE}$ from the last hidden layer.

$$
H^{VE}=\textrm{TransformerEncoder}(P)
$$

We utilize FastSpeech2's duration predictor to estimate the duration of each phoneme. This predictor's core component is a convolutional network \citep{lecun1998gradient} made up of two layers of one-dimensional convolutions. It takes the note information $U$, $V$, and the phonetic acoustic feature $H^{VE}$ as input.

$$
Q=\textrm{Convolution}(U\oplus V \oplus H^{VE})
$$

During training, we optimize the predictor by calculating the $L2$ norm of phoneme duration.
$$
\mathcal L_{dp}=-\mathbb{E}_{(Q)}(Q_i-\hat{Q}_i)^2
$$

The $F0$ predictor is based on a diffusion model \citep{nichol2021improved}. It first determines the total number of frames $T$ for the generated audio by adding up the phoneme durations $Q$ and creates a mapping matrix that connects phonemes to the mel spectrum. This mapping matrix is then used to extend the length of the acoustic features from $m$ to $T$. Finally, $H^{VE}_{1:T}$ is used as a condition and conduct $R^{'}$ rounds of denoising from the original noise $F0_{R^{'}}$ to reconstruct the pitch sequence. The $F0$ predictor is implemented using Wavenet \citep{oord2016wavenet}, which is a fully probabilistic and auto-regressive model composed of causal convolutional layers. During training, the $F0$ predictor employs the $L2$ norm as its loss function.

$$
F0=\textrm{Diffusion}(F0_{R^{'}},H^{VE}_{1:T},R^{'})
$$

After acquiring the phoneme duration $Q$ and $F0$, we will generate the Mel spectrum $M$ for the singing voice, which consists of $T$ frames. Initially, we will use an acoustic encoder to derive the acoustic representation $H^{AE}$. The inputs for the acoustic encoder include the phoneme $P$, phoneme duration $Q$, and $F0$, indicating that $H^{AE}$ captures essential information about the singing technique. The basic setup of the acoustic encoder mirrors that of the variance encoder.

$$
H^{AE}=\textrm{TransformerEncoder}(P\oplus Q\oplus F0)
$$

Subsequently, we will feed the vocal acoustic features $H^{AE}$ as a condition into a diffusion model to reconstruct the Mel spectrum of the singing voice. The standard inference process for the diffusion model is outlined, where $M_R$ represents the initial Gaussian noise and $R$ denotes the number of denoising iterations.
$$
M=\textrm{Diffusion}(M_R,H^{AE},R)
$$

To enhance the quality of the generated audio and speed up the inference of the diffusion model, we implement the shallow diffusion mechanism introduced by DiffSinger. This mechanism firstly employs a ConvNeXt-based auxiliary decoder \citep{liu2022convnet}, which consists of depth-wise convolutional blocks, to infer the spectrum $\tilde{M}_K$ at the $K$-th denoising step. 

$$
\tilde{M}_K=\textrm{Convolution}(H^{AE})
$$

Then, instead of conducting $R$ rounds of denoising from the original noise, the denoiser performs $K$ rounds starting from $\tilde{M}_K$.
$$
M=\textrm{Diffusion}(\tilde{M}_K,H^{AE},K)
$$

The denoiser's base model is identical to the $F0$ predictor. During the training phase, we enhance the auxiliary decoder by computing the $L1$ norm of the Mel spectrum, while the denoiser utilizes the $L2$ norm. 
$$
\mathcal L_{aux}=-\mathbb{E}_{(M)}|M_i-\hat{M}_i|
$$

The config file created in this section defines the singing style for the voice generation model and can also be used to alter the melody MIDI produced by the rhythm-to-melody module. Since the music generated by the model may not appeal to everyone, we believe that a music generation model allowing for flexible control over the output content is more universally applicable. Consequently, the config file generated by SongSong is accessible to users. If users are dissatisfied with the melody produced, they can modify the config file independently to adjust both the generated melody audio and voiceover audio at the same time.

\begin{figure}[htbp]
	\centering 
	\includegraphics[scale=0.6]{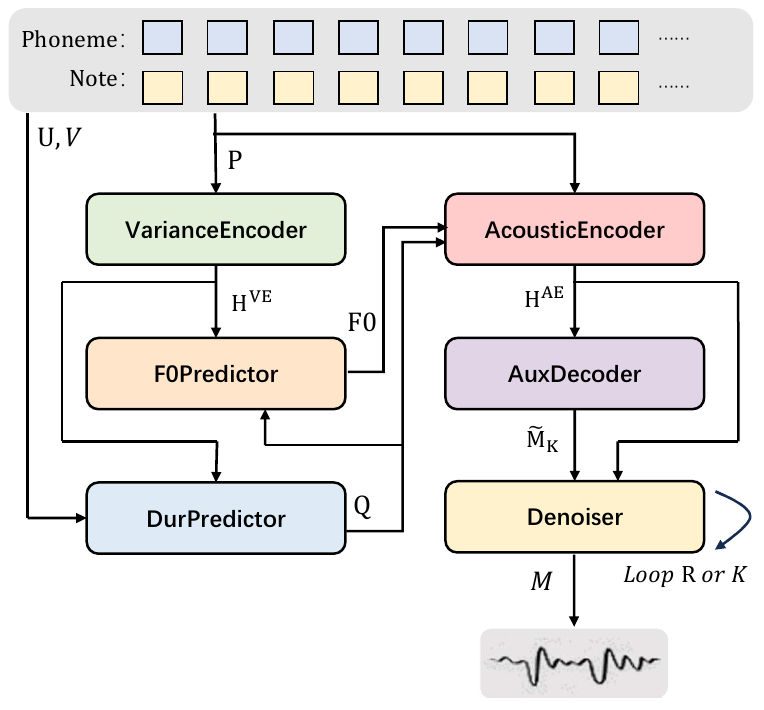} 
	\caption{The design of the singing voice generation module.}
    \label{singvoice}
\end{figure}

\subsection{Melody To Accompany}

After completing the previous two sections, we have acquired the vocal audio and melody audio, which can be merged to produce a full music track. To enrich the final music output, we can utilize an additional generative model to create accompaniments played by other instruments.
\begin{figure}[htbp]
	\centering 
	\includegraphics[scale=0.57]{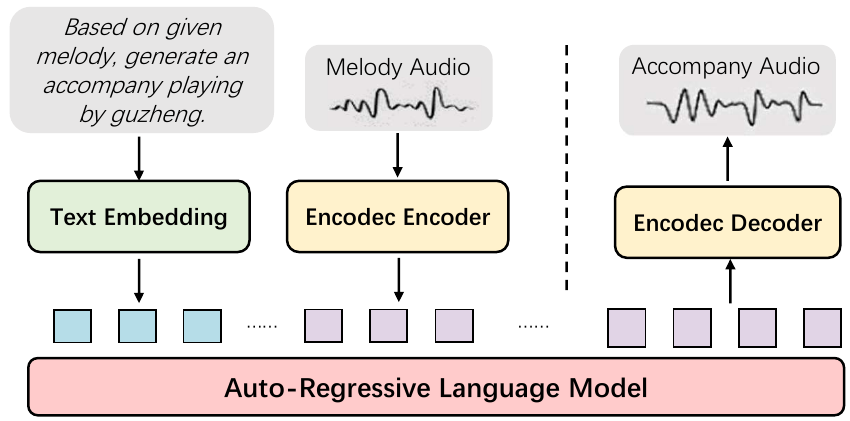} 
	\caption{The specific architecture of adopted accompany generation module.} 
    \label{accompany}
\end{figure}

\begin{table*}[htbp]
    \renewcommand{\arraystretch}{1.3} 
	\small  
		\begin{tabular}{ccccccccccc}	 
			\hline
			\multirow{2}{*}{Dataset}&\multirow{2}{*}{Style}&\multirow{2}{*}{Hours}&\multirow{2}{*}{Singers}&\multirow{2}{*}{Phoneme}&\multirow{2}{*}{Pitch}&\multirow{2}{*}{MIDI}&\multicolumn{4}{c}{Annotation}\\
            \cmidrule(lr){8-11}
            &&&&&&&Text&Phoneme&Pitch&Alignment\\
			\hline
			
			Opencpop &pop&5.25&1&miss ``van",``ve",``vn"&D\#3-D\#5&\ding{51}&\ding{51}&\ding{51}&\ding{51}&\ding{51} \\
            M4Singer&pop&29.77&20&no miss&G1-G5&\ding{51}&\ding{51}&\ding{51}&\ding{53}&\ding{51} \\
			\hline
            \textbf{OpenSongSong}&SongCi&29.90&89&no miss&G1-C6&\ding{51}&\ding{51}&\ding{51}&\ding{51}&\ding{51} \\
            \hline
		\end{tabular}
	\centering
	\caption{The comparison results for OpenSongSong, Opencpop, and M4Singer. The ``Phenome" indicates if certain phonemes are absent from the dataset, while the ``Pitch" reflects the distribution of pitches within the dataset. Phonemes or pitches that occur fewer than 20 times are deemed excluded from the dataset. The ``MIDI" indicates whether the dataset includes music score information. The ``Annotation" specifies the number of annotations present in the dataset. Lastly, the ``Alignment" indicates whether each annotation is synchronized with the corresponding singing voice recording, allowing for the start and end times of each element to be extracted from the annotation type.}
    \label{datasets}
\end{table*}

Following the approach outlined by \citet{copet2024simple}, we implement an architecture that discretizes acoustic features and generates discrete acoustic tokens through an autor-egressive language model for accompaniment generation. Figure \ref{accompany} illustrates the specific structure of the accompaniment generation module. The input for this module consists of a text prompt and an audio prompt, with the audio prompt being the melody audio. The continuous audio features are quantized using the encoder from Encodec \citep{defossez2022high}. During inference, the text tokens and audio tokens are combined and fed into an auto-regressive Transformer to produce the accompaniment audio tokens. This tokens are then transformed into a continuous Mel spectrogram using the decoder from Encodec.

\section{Experiment}
\subsection{Dataset}

OpenSongSong is an extensive collection of Chinese SongCi music that includes recordings of singing voices about 29.9 hours along with their associated annotations and musical score details. The annotations are stored in a textgrid file, which contains text, phoneme sequences, and pitch sequences. Each annotation is synchronized with the singing voice audio, allowing for easy access to the start and end times of each element directly from the textgrid file. The musical score information is converted into MIDI format, which represents a musical composition in written form, typically detailing note pitch, duration, and tempo. 



We compare the richness of Opencpop and M4Singer with OpenSongSong in Table \ref{datasets}. The size of OpenSongSong is comparable to that of M4Singer, and both datasets encompass all Chinese phonemes. However, OpenSongSong features a greater number of singers and a broader pitch range. Additionally, both OpenSongSong and Opencpop include pitch information in the textgrid file, eliminating the need to extract pitch from the audio again, thus facilitating its use in SVS tasks. OpenSongSong stands as the first large-scale, high-quality dataset of SongCi music, offering diverse annotations, covering all Chinese phonemes, and supporting a wide pitch distribution. It is suitable for various music-related applications, including singing voice generation, accompaniment generation, and lyrics-melody alignment.

\begin{table*}[htbp]
    \renewcommand{\arraystretch}{1.3} 
    \setlength{\tabcolsep}{5pt} 
	\small 
		\begin{tabular}{cccccccccccc}	 
			\hline
			\multirow{2}{*}{Test}&\multirow{2}{*}{Model}&\multicolumn{2}{c}{Objective Metrics}&\multicolumn{4}{c}{Common Subjective Metrics}&\multicolumn{4}{c}{SongCi Subjective Metrics} \\
            \cmidrule(lr){3-4}\cmidrule(lr){5-8}\cmidrule(lr){9-12}
            &&$\mathrm{FAD_{vgg}}$ ($\downarrow$) &$\mathrm{FAD_{pann}}$ ($\downarrow$)&$\mathrm{MS}$ ($\uparrow$) &$\mathrm{ER}$ ($\uparrow$) & $\mathrm{MC}$ ($\uparrow$) &$\mathrm{SQ}$ ($\uparrow$)& $\mathrm{\textbf{PA}}$ ($\uparrow$)& $\mathrm{\textbf{VN}}$ ($\uparrow$) & $\mathrm{\textbf{SCS}}$ ($\uparrow$)& $\mathrm{\textbf{AC}}$ ($\uparrow$)\\
			\hline
            \multirow{3}{*}{Zero-shot}&Suno&5.74&\textbf{3.35e-4}&65.48&68.85&63.19&62.96&62.59&65.37&35.30&19.28\\ 
            &SkyMusic&7.88&3.42e-4&63.96&65.48&59.11&72.67&54.81&\textbf{72.89}&25.37&45.69\\
            \cmidrule{2-12}
            &SongSong&\textbf{5.41}&3.87e-4&56.67&46.26&53.93&55.56&\textbf{75.37}&65.56&\textbf{78.44}&\textbf{65.54}\\
            \hline
            \multirow{3}{*}{Few-shot}&Suno&7.92&3.98e-4&64.44&62.30&64.41&75.15&48.85&71.04&42.19&28.90\\
            &SkyMusic&6.31&\textbf{5.55e-5}&60.26&58.30&60.63&76.41&59.04&72.74&46.67&52.60\\
            \cmidrule{2-12}
            &SongSong&\textbf{5.41}&3.87e-4&50.52&49.48&50.04&63.74&\textbf{78.48}&\textbf{72.81}&\textbf{75.19}&\textbf{69.76}\\
            \hline
		\end{tabular}
	\centering
	\caption{The comparison results for SongSong, Suno, and SkyMusic. The terms zero-shot and few-shot indicate whether Songci audio was provided as a reference for Suno and SkyMusic. For subjective metrics, $\mathrm{MS}$ stands for music structure, $\mathrm{ER}$ denotes equipment richness, $\mathrm{MC}$ indicates motivation continuity, $\mathrm{PA}$ refers to pronunciation accuracy, $\mathrm{VN}$ signifies voice naturality, $\mathrm{SQ}$ represents sound quality, $\mathrm{SCS}$ pertains to conformity with the SongCi style and $\mathrm{AC}$ is acompaniment conformity.}
    \label{mainresult}
\end{table*}

\subsection{Configuration}

Both the lyric-to-rhythme and rhythm-to-melody modules share the same Transformer architecture, consisting of 4 layers of encoder blocks and 4 layers of decoder blocks. Each block features 4 attention heads with 256 linear units. In the singing voice generation module, our acoustic encoder and variance encoder are based on FastSpeech2, following the same configuration as \citet{ren2020fastspeech}, which includes 4 feed-forward Transformers. The duration predictor comprises 5 layers of convolutional networks, each with a one-dimensional convolution and a kernel size of 3, maintaining an input and output size of 512. The auxiliary decoder consists of 6 ConvNeXt decoder layers, each with a kernel size of 7 and an input and output size of 512. The denoiser and F0 predictor are both based on Wavenet, featuring 20 convolutional layers with a kernel size of 3. The input and output sizes for the denoiser in each layer are 512 and 1024, respectively, while the F0 predictor has half the size. For the accompaniment generation module, we utilize 1.5B-Musicgen-melody as the initial parameters.

We utilize data from the speaker with the largest corpus from OpenSongSong, which amounts to around 3.5 hours, to train the lyrics-to-rhythm, rhythm-to-melody, and singing voice generation components of SongSong. The rhythm and melody corpora are prepared following the methodology outlined by \citet{ju2022telemelody}. All training sessions consistently use the Adam optimizer with a learning rate set at 5e-4. The training is performed on GeForce RTX-3090, with all modules trained for a maximum of 160,000 steps.


%

Currently, there are limited open-source systems capable of generating audio that combines both singing and accompaniment. Therefore, we select Suno and SkyMusic, two leading commercial music generation software, for comparison. These systems leverage the advanced generation capabilities of LLMs to produce high-quality music that includes both singing and accompaniment based on simple text prompts. Additionally, they allow users to input an audio reference to generate music in a similar style.



We use objective and subjective metrics for evaluation. The objective metric employed is the Frechet Audio Distance \citep{kilgour2018fr}, which measures the Frechet distance between the embeddings from two audio groups, providing an evaluation of generated audio quality that aligns more closely with human perception. We utilize the VGGish and PANN \citep{kong2020panns} models to extract audio embeddings. For subjective evaluation, we evaluate the music structure, richness of instrumentation, continuity of motivation, sound quality, pronunciation accuracy, naturalness of voice, adherence to the SongCi style and accompaniment conformity. The first four metrics are evaluated from the perspective of ordinary music, while the last four metrics evaluate the generated music from the perspective of SongCi music.

We randomly select 50 songs that are not part of the training data from the SongCi collection used to develop OpenSongSong for our test set. After performing basic segmentation to eliminate long stretches of silence, we end up with 85 utterances totaling 1.8 hours. The singer features in this test set is male. The testing process consists of two phases. The first phase is a zero-shot test, where only text prompt and SongCi lyric are provided to Suno and SkyMusic, with the prompt specified as “Chinese classical style, Song Dynasty, Guzheng, male voice.” The second phase is a few-shot test, where Suno and SkyMusic receive an additional SongCi piece as a reference. We invite 9 experts from music academies to participate in the subjective evaluation.


\subsection{Main Result}

\begin{figure*}[htbp]
	\centering 
	\includegraphics[scale=0.43]{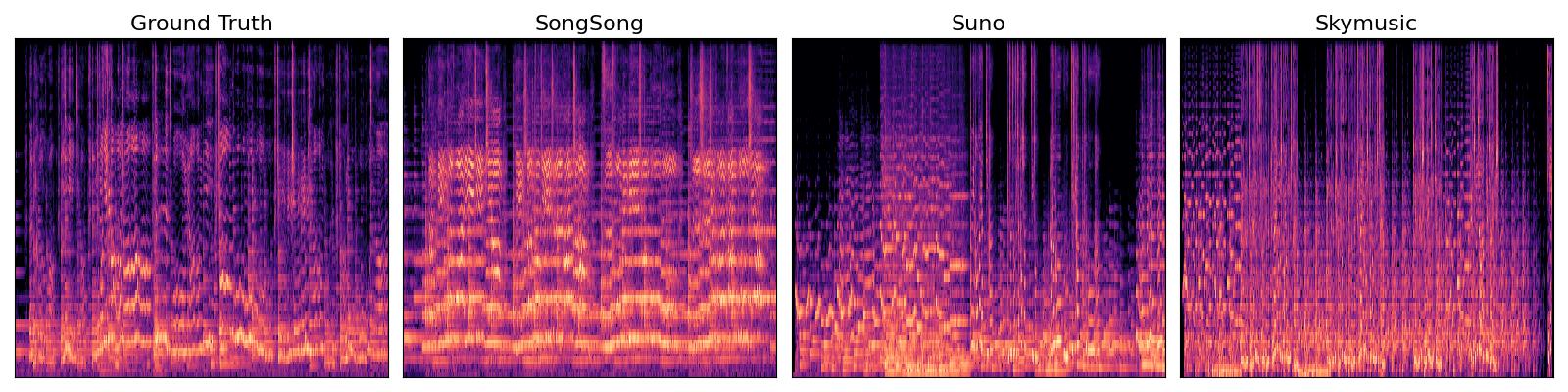} 
	\caption{The mel-spectrogram comparisons among ground truth, SongSong, Suno and Skymusic.} 
    \label{casestudy}
\end{figure*}

The experimental results are shown in Table \ref{mainresult}. When conducting subjective evaluations from the perspective of SongCi music, we find that Suno and SkyMusic have poor prompt acceptance ability and cannot understand Chinese classical styles well. The generated music is almost all in the style of pop music, and the musical characteristics are far from that of SongCi music. Moreover, the musical accompaniment does not use the guzheng as we have defined, but uses instruments such as guitars and keyboard instruments that are typically used in pop music. This problem is difficult to alleviate even when inputting SongCi music as a reference. Therefore, Suno and SkyMusic score much lower than SongSong on SCS and AC. SongSong, after being trained on the SongCi musical corpus in OpenSongSong, can fully grasp the pitch and rhythm of SongCi music, and generate singing voice and accompaniment that are more in line with SongCi music. In addition, due to the presence of many rare characters in Song poetry, the two large music models trained on popular music cannot clearly and accurately sing every word, and sometimes even miss words and generate random lyrics. This results in their PA being low. After few-shot, the accuracy of SkyMusic's pronunciation shows an upward trend, while that of Suno shows a downward trend. However, this issue has not significantly affected the VN of Suno and SkyMusic. With the large scale pre-training and a large number of model parameters, Suno and SkyMusic can still maintain the naturalness of the original performance to a certain extent, so VN is not lower than SongSong. SongSong has undergone strict alignment of lyrics and melody, and the training corpus already includes rare words that may appear in SongCi music, so it can not only sing strictly according to the input lyrics, but also sing each word with correct pronunciation. Therefore, it far exceeds the other two models in the PA metric and is comparable to them in VN. However, when evaluating subjectively from the perspective of ordinary music, SongSong's performance is not as good as Suno and SkyMusic. In addition to the backwardness in model size and training data capacity, there are several reasons for this phenomenon. In a SongCi poem, the sentence structure and length are relatively fixed, and the performance style is relatively monotonous. But Suno and SkyMusic sing it as a popular song, using structural optimization methods such as repetition, contrast, and modulation to generate more complex and continuous pop music, thus having higher MS and MC; The controllability of Suno and SkyMusic is poor. Although we input a prompt requiring them to only use the guzheng for accompaniment, they still use other instruments for accompaniment, resulting in a higher degree of orchestration, so their ERs are high; After few-shot, MS and ER decrease for Suno and SkyMusic, while MC remains unchanged. This observation supports our analysis that the closer the generated music is to SongCi, the more monotonous and consistent its style will be. In terms of sound quality, due to the lack of high-quality Song Dynasty music, our trained model can not generate muisc with very high sound quality. In terms of objective evaluation metrics, SongSong has a relatively high $\mathrm{FAD_{VGG}}$, but it is comparable to Suno and SkyMusic in terms of $\mathrm{FAD_{PASS}}$. The audio for comparison is SongCi audio, so objectively speaking, SongSong can also generate high-quality SongCi audio that conforms to human auditory perception.

\subsection{Case Study}

To showcase SongSong's capability to produce high-quality audio that aligns more closely with the style of SongCi music, we carry out a case study using a sample from a test set. Figure \ref{casestudy} displays the mel-spectrogram of the sample alongside the audio generated by SongSong, Suno, and SkyMusic based on the sample's lyrics. The original mel-spectrogram is repetitive over time due to the consistent sentence structure of the SongCi and the fixed rhythm of the SongCi music. In contrast, Suno's generated mel-spectrogram exhibits significant variation in frequency and loudness across different time intervals, suggesting a high level of randomness which shows Suno fails to capture the monotonous and repetitive nature of Song poetry. Additionally, it lacks mid-frequency and high-frequency bands. SkyMusic performs better than Suno in generating Chinese SongCi music, with its mel-spectrogram reflecting the overall style of Song poetry; however, it suffers from abrupt changes in the high-frequency range. On the other hand, the mel-spectrogram produced by SongSong shows clear temporal repetition, indicating that it has effectively learned the principles of SongCi music and can maintain a consistent rhythm for each SongCi line, performing them in accordance with the SongCi style.

\subsection{Ablation Study}
\begin{table}[htbp]
    \renewcommand{\arraystretch}{1.3} 
    \setlength{\tabcolsep}{1.5pt} 
	\small  
		\begin{tabular}{cccccc}
			\hline
			\multirow{2}{*}{Model}&\multicolumn{2}{c}{Objective Metrics}&\multicolumn{3}{c}{SongCi Subjective Metrics} \\
            \cmidrule(lr){2-3}\cmidrule(lr){4-6}
            &$\mathrm{FAD_{vgg}}$ ($\downarrow$) &$\mathrm{FAD_{pann}}$ ($\downarrow$) &$\mathrm{PA}$ ($\uparrow$)& $\mathrm{VN}$ ($\uparrow$) & $\mathrm{SCS}$ ($\uparrow$) \\
			\hline
            GSV&20.30&\textbf{4.93e-4}&45.25&69.25&46.25 \\
            SongSong&\textbf{7.22}&5.05e-4&\textbf{81.25}&\textbf{70.00}&\textbf{75.50} \\
            \hline
		\end{tabular}
	\centering
	\caption{The results of ablation study.}
    \label{ablation}
\end{table}
In essence, Suno adopts a structure similar to VALL-E \citep{wang2023neural}, discretizing acoustic features into tokens, and then leveraging LLM to sequentially decode the acoustic tokens to generate a complete piece of music. But in fact, this method is not suitable for low-resource situations, where the melody generation module plays a key role. To prove this, we additionally introduce GPT-SoVITS (GSV), a SOTA speech synthesis model, which is based on \citet{brown2020language} and \citet{kim2021conditional}, has the above architecture. It can be trained using the same training data as SongSong to obtain the ability to generate singing voice. During the evaluation, we select 20 unused SongCi audio clips and their corresponding lyrics, and ask GSV and SongSong to generate SongCi music based on the lyrics. GSV uses a piece of SongCi music for few-shot. We assess the FAD based on the original SongCi audio clips, and we also invite two music experts to evaluate the performance of the generated audio on subjective metrics of SongCi music. The experimental results are shown in Table \ref{ablation}. In terms of objective metrics, SongSong significantly outperforms GSV on $\mathrm{FAD_{vgg}}$. For three subjective metrics of SongCi, GSV's performance is also far inferior to SongSong. During the subjective evaluation, our music experts point out that although GSV can sing SongCi with a certain rhythm, it cannot completely sing the input SongCi, and there are serious phenomena of missing words and repeated generation. Even though GSV has been trained on SongCi music data and uses few-shot in inference, it is difficult to align lyrics and melody due to limited training corpus, so it cannot strictly output based on the input SongCi. SongSong uses a rhythm-based lyric-melody alignment scheme to generate melody information first, and then uses both the lyric information and the melody information to generate singing voice, which is a high-quality audio that can fully perform the input lyrics and conform to the style of SongCi.


\section{Conclusion}
We present the first music generation model capable of performing Chinese SongCi, SongSong, and the first comprehensive dataset of ancient Chinese SongCi music, OpenSongSong, which supports the restoration of SongCi music. SongSong first predicts the melody from the input SongCi, then separately generates the singing voice and accompaniment based on that melody, and finally combines all audio elements to create the final SongCi music. OpenSongSong features 29.9 hours of SongCi music. We evaluate the performance of SongSong, Suno and SkyMusic, and the results indicate that our proposed model can produce high-quality music that embodies the SongCi style.

\section*{Acknowledgements}
This work was supported by the National Natural Science Foundation of China (No. 62306216, No. 72074171, No. 72374161), the Natural Science Foundation of Hubei Province of China (No. 2023AFB816), the Fundamental Research Funds for the Central Universities (No. 2042023kf0133). 

\bibliography{custom}

\end{document}